# Electron gun system for NSC KIPT linac[1*]

ZHOU Zu-sheng[1)], HE Da-yong, CHI Yun-long

Institute of High Energy Physics, CAS, Beijing 100049, China

**Abstract** In NSC KIPT linac, a neutron source based on a subcritical assembly driven by a 100MeV/100kW electron linear accelerator is under design and development. The linear accelerator needs a new high current electron gun. In this paper, the physical design, mechanical fabrication and beam test of this new electron gun are described. The emission current is designed to be higher than 2A for the pulse width of 3us with repetition rate of 50 Hz. The gun will operate with a DC high voltage power supply which can provide a high voltage up to 150 kV. . Computer simulations and optimizations have been carried out in the design stage, including the gun geometry and beam transport line. The test results of high voltage conditioning and beam test are presented. The operation status of the electron gun system is also included. The basic test results show that the design, manufacture and operation of the new electron system are basically successful.

## 1    Introduction

In the NSC KIPT linac (National Science Center, Kharkov Institute of Physics and Technology,) a neutron source based on a subcritical assembly driven by a 100MeV/100kW electron linear accelerator is under design and development [1]. The linear accelerator needs a new high current electron gun. The electron gun is a conventional thermionic triode electron gun which consists of an anode, a cathode and a grid. It is equipped with Y824 (EIMAC) cathode grid assembly. The electron gun system consists of an electron gun body, a high voltage power system, a high voltage deck, a pulser, a filament power and a control unit, etc. The gun can operate in long pulse mode to generate low emittance beam. The main parameters of the electron gun are shown in Table 1[2].

Table1    The electron gun parameters

| Item | Specification | Unit |
|---|---|---|
| Type | Triode | |
| Beam Current（max） | 2 | A |
| Anode Voltage | ~120 | kV |
| Filament Voltage | 6.4 | V |
| Filament Current | 5.5 | A |
| Grid Bias | 50~ 500 | V |
| Bunch Length | 3.0 | μs |
| Repetition Rate (max) | 625 | Hz |

## 2   Gun design consideration

### 2.1   Gun body

A planar triode electron gun, the EIMAC Y824 with 150 kV GLASSMAN high-voltage deck, is used with a fast pulser cathode driver. The acceleration voltage is 120 kV, the maximum extraction current is 2A (normal operation is about 0.85A) with minimum pulse width of less than

[1]*Supported by National Natural Science Foundation of China (11075172, 11075173)

1) E-mail: zhouzs@ihep.ac.cn

3.0μs. The gun pulser system contains a driving circuit. A 3.0μs FWHM pulser is used to drive the cathode for long pulse operation. The gun structure and the gun beam pulse form are shown in Fig.1.

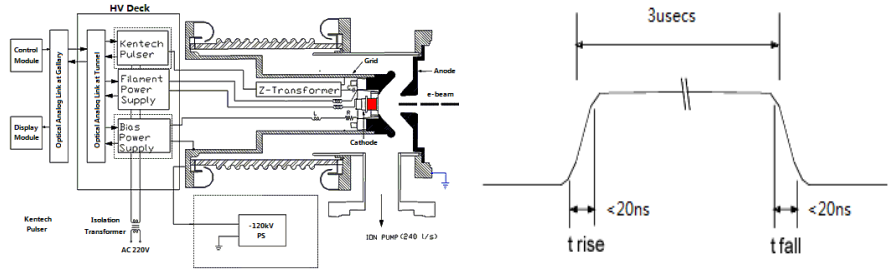

Fig.1 The gun structure (left) and the gun beam pulse form (right)

## 2.2　Beam optic simulation

In order to achieve the design goal, it is necessary to use an electron gun with good performance and high emission current. EGUN[3] program is used to optimize the shapes and dimensions of the focusing electrode and anode.

The beam simulation result is shown in Fig.2. The simulation results show that if the beam current is about 2A at high voltage of 120kV for example, then the beam emittance is about 5.7π·mm·mrad, which can easily satisfy the requirements of accelerator with normal operation current of 0.85A[4].

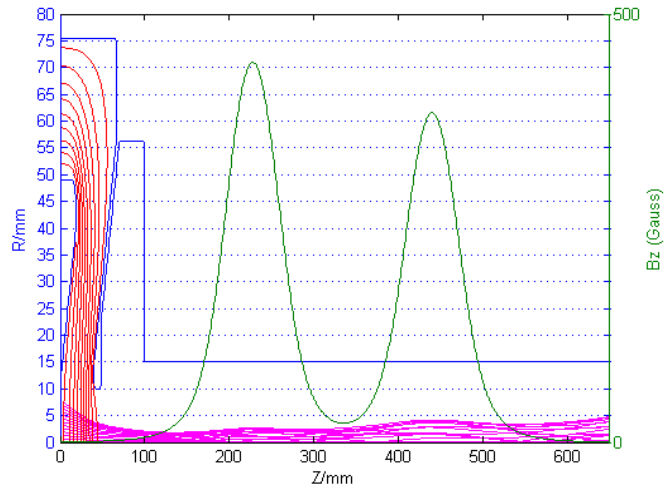

Fig.2 Beam optics with 2.0A/120kV in the gun

## 2.3　Beam transport and elements

Two magnetic lenses and two sets of steering coils are adopted in our design to focus and adjust the beam between the gun and the bunching system. A few beam instrumentation elements, such as beam position monitors and beam profile monitor, are placed between the gun and the bunching system. With such instruments, we can tune the beam flexibly and reliably. After compromising with some installation problems, the element distribution is shown in Fig.3.

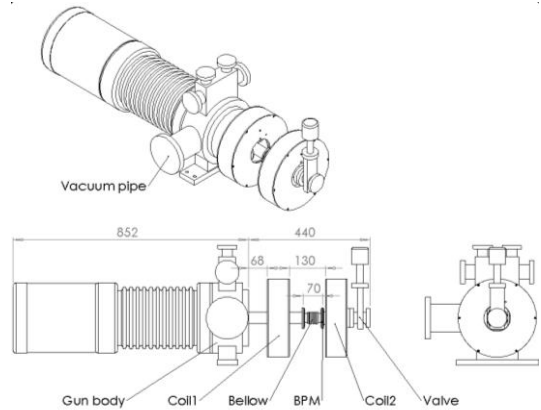

Fig.3 Element distributions between the gun and the bunching system

## 3    Gun conditioning and test

After the arrival of the electron gun body from the machine shop, the new electron gun is tested on the test bench. These include the building of an experimental platform, the installing of the electron gun, the internal vacuumizing of the gun, the cathode grid assembly activation experiment, and finally the electron gun high voltage conditioning and beam test.

### 3.1   Installation

After installing the electron gun body, the cathode grid assembly is installed, and the vacuum leak detector shows that the background has no leakage and the vacuum inside the gun meets the test requirements. Fig.4 shows the gun body on the test stand and the cathode grid assembly installation.

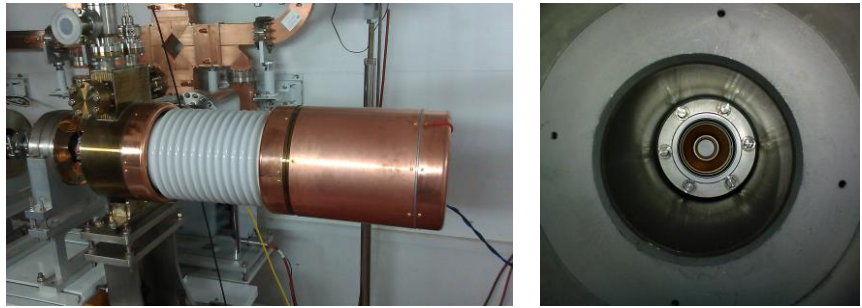

Fig.4 The gun test stand (left) and cathode grid assembly installation (right)

### 3.2   High voltage conditioning

Before the electron gun beam test, the gun body needs to have high voltage conditioned between the cathode and the anode through the multiple low energy discharge between the electrodes. The conditioning process removes contaminants on the insulating ceramic and the electrode surface, eliminates burrs and micro-protrusion of the surface of the electrode, and improves the breakdown voltage of the electrodes.

Fig.5 shows the sketch map of high voltage conditioning for electron gun. The capacity of energy storage $C_t$ is 4920pF, it stores 35.4J at 120kV, when the gun body arcs down to the earth at 120kV the energy will be absorbed by 10 resistors in series whose rating power is 300W. The pulse width t is $3 \mu S$, the pulse current $I_p$ is 2A, then the flatdrop of pulse $\Delta U$ is 1219V. The percentage of flatdrop is 1.0%. Filament power supply and pulse power supply are provided by AC220V power on the HV platform isolated by a 150kV transformer. The maximum voltage is about 135 kV on electron gun given by H.V power supply.

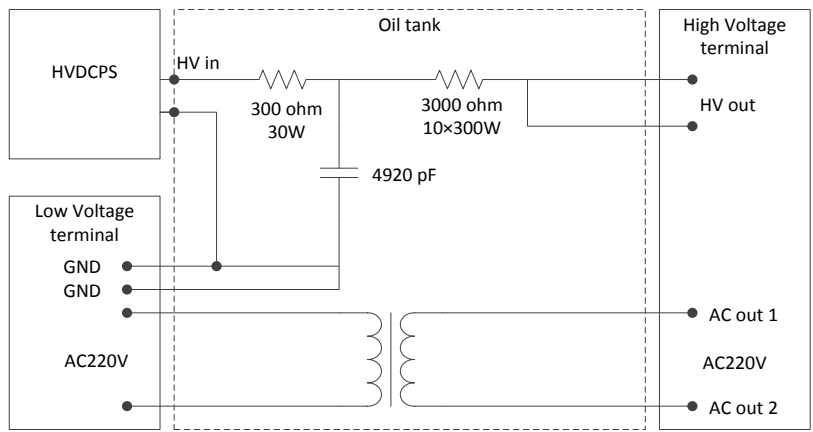

Fig.5 High voltage conditioning sketch map

### 3.3  Cathode activation

Activation is achieved by converting the barium oxide in the tungsten matrix into free barium on the surface of the cathode. The rate of activation is a function of tube cleanliness, cathode poisoning, time and temperature. In our system, the maximum power for filament is about 8.1V/6.3A (keeping 5 min.); the operating power is about 6.4V/5.5A. The activation time is up to the vacuum system and keeps the pressure at $1 \times 10^{-6}$ Torr or better.

### 3.4 Beam test

One Faraday cup is mounted after the anode of the electron gun, and an oscilloscope is connected to it. With 40dB attenuation, the Faraday cup signal is directly given by the oscilloscope. The Faraday cup has an impedance of 50 Ohm, and the oscilloscope's input impedance is also 50 Ohm. We have the following approximate relationship: the beam current equals the oscilloscope voltage readings plus 100/50, i.e the beam current equals the oscilloscope voltage readings plus 2.

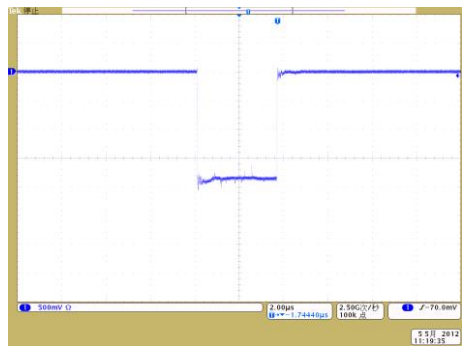

Fig.6 The gun current signal from oscilloscope

3.5A electron beam was successfully obtained when we applied 120kV high voltage at repetition rate 1~10Hz, the flat top width of the pulsed power supply output is about 3.5us as shown in Fig. 6[5].

The typical gun characteristics are shown in Fig. 7, which shows the gun current as a function of gun high voltage under different heater power; and Fig.8 shows the gun current as a function of grid bias voltage. In Fig. 7, grid bias voltage is set to a fixed value, -100V; and the pulser output is also fixed, about 200V. In Fig. 8, the filament power is fixed to about 5.8A, and the anode high voltage relative to anode is 120 kV.

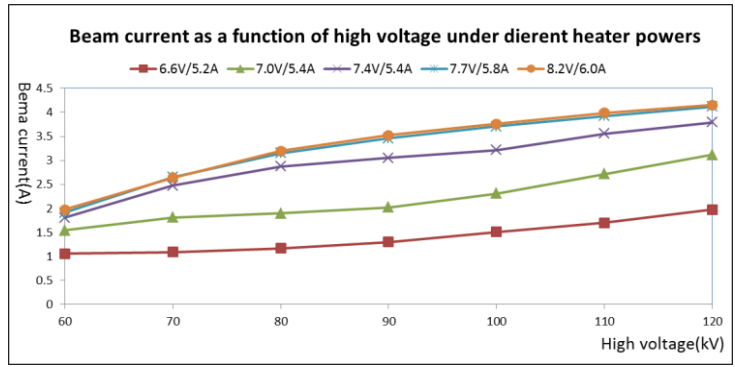

Fig.7 Beam current vs. high voltage

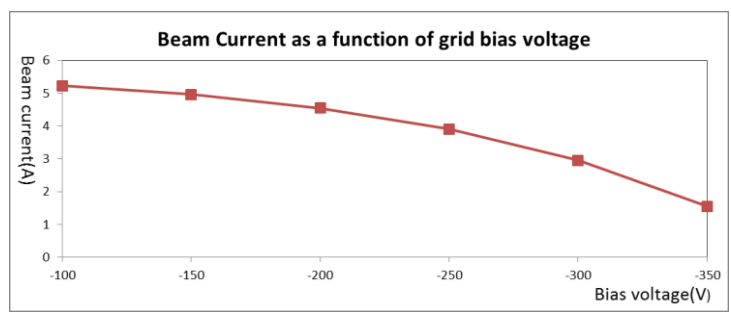

Fig.8   Beam current vs. grid bias voltage

## 4    Conclusions

The new electron gun system for the NSC KIPT linac is now operated under 120 kV with repetition rate of 1~50Hz. The maximum beam current is about 3.5A at the gun exit. The beam pulse length is about 3.5us, which is enough for the linac operation. At this moment, the basic test results show that the design, manufacture and operation of the new electron system are basically successful. Moreover, there is the first China-made pulser system used in the new electron gun system and the new ceramic seal structure is available for the spare electron gun system.